\documentclass[12pt]{iopart}

\usepackage{iopams}  

\newtheorem{theorem}{Theorem}[section]

\usepackage{graphicx,color}
\usepackage{amstext}
\usepackage{hyperref}
\usepackage{mathrsfs}

\def\e{\mathrm{e}}
\def\ii{\mathrm{i}}
\def\d{\mathrm{d}}

\def\R{\mathbb{R}}

\def\I{I}

\begin{document}
\title[]{Quantum cavities with alternating boundary conditions}
\author{Paolo Facchi$^{1,2}$, Giancarlo Garnero$^{1,2}$ and Marilena Ligab\`o$^{3}$}
\address{$^{1}$Dipartimento di Fisica and MECENAS, Universit\`a di Bari, I-70126  Bari, Italy}
\address{$^{2}$INFN, Sezione di Bari, I-70126 Bari, Italy}
\address{$^{3}$Dipartimento di Matematica, Universit\`a di Bari, I-70125  Bari, Italy}
\date{\today}

\begin{abstract}
We consider the quantum dynamics of a free nonrelativistic  particle moving in a cavity and we analyze the effect of a rapid switching between two different boundary conditions. We show that this procedure induces, in the limit of infinitely frequent switchings, a new effective dynamics in the cavity related to a  novel boundary condition. 
We obtain a dynamical composition law for boundary conditions which gives the emerging boundary condition in terms of the two initial ones.

\end{abstract}


\vspace{2pc}

\maketitle

\section{Introduction} \label{sec-dpw}
Boundary conditions emerge as an effective description of the interaction between confined physical systems and their environment. 
In quantum theory, observables in physical systems are associated with self-adjoint operators~\cite{von}. In some cases, physical reasoning gives a formal expression for the Hamiltonian operator on a given Hilbert space. Such an operator is in general unbounded and, for this reason, it is necessary to consider the domain where it is defined. 

Usually, one starts by considering some natural domain, where the operator is symmetric, and tries to verify whether the operator is self-adjoint or not. In the negative case one is forced to study various (if any) self-adjoint extensions of the operator under examination. Moreover, in the spirit of Stone's theorem~\cite{rs}, self-adjoint operators can be interpreted as the  generators of the physical (unitary) transformations on the Hilbert space of physical states, so that, the choice of the ``right'' self-adjoint extension is usually guided by physical intuition.
For example,  a free quantum particle confined in an impenetrable box is usually described in terms of the Laplace operator with Dirichlet boundary conditions. In this case it is necessary to impose the vanishing of the wave function at the boundary of the box in order to secure the unitarity of the dynamics.

Many authors have studied the connection between self-adjoint extensions and boundary conditions, see for example~\cite{grubb68,da,p,aim,extensions}, and the latter have been proved to be useful in the descriptions of quantum phenomena in different areas of physics, ranging from atoms in cavities~\cite{HarocheRaimond}, to edge states in the quantum Hall effect~\cite{abp}, to quantum gravity and string theory~\cite{Wilczek}. 

Moreover, it is well known that the introduction of boundaries in quantum systems gives rise to peculiar phenomena like the Casimir effect. In its dynamical counterpart, photons can be generated in a microwave cavity as long as time-dependent boundary conditions are implemented~\cite{photons}.

Time-dependent boundary conditions~\cite{moving walls,bangalectures} can also produce a non-trivial Berry phase, as shown in~\cite{pgbj}. Interestingly the simplest model of hyperbolic geometry, the Poincar\'e half-plane, emerges, bringing to light an unexpected link between geometry and quantum boundary conditions.

In this paper  we will consider one of the paradigmatic examples of nonrelativistic quantum mechanics, that is, the case of a free particle confined in a cavity, i.e.\ in a region $\Omega$ of the $n$-dimensional space $\mathbb{R}^n$. The cavity is going to be subjected to a rapid switching between two different boundary conditions. We are going to show that the emerging dynamics, in the limit of infinitely frequent switchings, yields new boundary conditions obtained by suitably combining the initial ones.

This work represents the multidimensional generalization of the results in~\cite{composition1}, where the one-dimensional case, i.e.\ a free particle confined on an interval, is studied. The  crucial difference between the one-dimensional and the higher-dimensional case is the dimension of the boundary. If the particle is confined in a region $\Omega \subset \mathbb{R}^n$, the boundary $\partial \Omega$ is  a manifold of dimension $n-1$. Thus, if  $n=1$  and $\Omega$ is an interval its boundary has dimension zero and consists of two points. This means that there are only two boundary values for the wave function and other two for its derivative. In the  $n$-dimensional case, with $n>1$, the boundary values of a wave function are functions defined on $\partial \Omega$ and this aspect makes the extension to the general case very delicate. Indeed, for $n>1$ the Hilbert space on the boundary is infinite dimensional, at variance with a finite-dimensional Hilbert space for $n=1$. This entails the emergence of qualitatively new and interesting phenomena, which were not present in the one-dimensional case.

The article is organized as follows. In Section~\ref{sec-trotter} we introduce the problem of composition law for boundary conditions and, after a short review of standard results about product formulas, we present our  main result. In Section~\ref{sec-interval}, we revise the composition law for one-dimensional systems, discussed in~\cite{composition1}, in terms of a different approach, which will turn out to be  suitable for the extension to the $n$-dimensional case. Finally, in Section~\ref{sec-cavity} some results on self-adjoint extensions are recalled and used for obtaining the composition law for a particle confined in an $n$-dimensional cavity with alternating boundary conditions. We finally draw our conclusions in Section~\ref{sec-concl}. 
In two appendices we also provide some technical results which turn out to be  useful in Section~\ref{sec-cavity} to define the boundary data of a wave function in the $n$-dimensional case.

\section{Trotter formula for alternating boundary conditions} \label{sec-trotter}
Let us consider a quantum particle confined in a cavity $\Omega\subset\mathbb{R}^n$, where $\Omega$ is an open connected set, whose boundary $\partial\Omega$ is a smooth submanifold of $\mathbb{R}^n$.
We suppose that the particle is spinless, free and has a mass $m$. The description of this system is provided by the free Hamiltonian 
\begin{equation}
T=\frac{p^2}{2m}= -\frac{1}{2m}\Delta,
\end{equation}
that is the the kinetic energy operator (here and henceforth, Planck's constant  $\hbar=1$). Here $\Delta$ is the Laplacian defined on some dense subspace of $L^2(\Omega)$, the Hilbert space of square integrable functions on $\Omega$.

We are going to denote by $\mathscr{H}_b$ the Hilbert space of the boundary values of all possible wave functions, which will be specified later. It can happen that the boundary value of a function in $L^2(\Omega)$ is a distribution (see the example at the end of~\ref{sec:appendixA}). For this reason, as we are going to discuss later on, $\mathscr{H}_b$ cannot be identified with the space of square integrable functions on the boundary, $L^2(\partial\Omega)$. 

It was proved in~\cite{aim,extensions}, that every unitary operator $U$ acting on $\mathscr{H}_b$  uniquely determines a well-defined physical realization of $T$, i.e a self-adjoint extension of $T$, denoted by $T_U$, identifying some boundary conditions. The  relation between the self-adjoint extension $T_U$ and the unitary operator $U$ will be recalled in Sections~\ref{sec-interval} and~\ref{sec-cavity}.

Now,  consider $T_{U_1}$ and $T_{U_2}$, two realizations of $T$, corresponding to two boundary conditions, given by $U_1$ and $U_2$, unitary operators on $\mathscr{H}_b$. Moreover we let the two boundary conditions rapidly alternate.

More precisely, if the boundary conditions are switched at a rate $t/N$, the overall unitary evolution reads
\begin{eqnarray}
\label{eq:2evol}
& & \underbrace{\left(\e^{-\ii t T_{U_1}/N} \e^{-\ii t T_{U_2}/N}\right)
\left(\e^{-\ii t T_{U_1}/N} \e^{-\ii t T_{U_2}/N}\right)
\dots
\left(\e^{-\ii t T_{U_1}/N} \e^{-\ii t T_{U_2}/N}\right)}_{N \; \mathrm{ times}} \nonumber \\
& & \quad =\left(\e^{-\ii t T_{U_1}/N} \e^{-\ii t T_{U_2}/N}\right)^N ,
\end{eqnarray}
Equation~(\ref{eq:2evol}) is an example of Trotter's product formula~\cite{Trotter}. Similar evolutions show up in different contexts of quantum mechanics, for example, in the emerging dynamics associated with the Quantum Zeno effect~\cite{exnerrev,fl,graffi,Zenolimit}, as well as in remarkable applications in quantum chaos~\cite{qchaos1,qchaos2}. Lastly, but not less importantly, the Trotter product formula is fundamental in a rigorous definition of  Feynman's path integral~\cite{Lapidus}.

As already discussed in~\cite{composition1}, the relevant question is to show whether in the $N \to + \infty$ limit---when the time interval between the switches goes to zero, the number of switches goes to infinite, while the total time $2t$ is kept constant---the evolution is given by
\begin{equation}
\label{eq:evollim}
\left(\e^{-\ii t T_{U_1}/N} \e^{-\ii  t T_{U_2}/N}\right)^N \to
\e^{-\ii 2 t T_{W}}, \quad N \to +\infty
\end{equation}
in terms of a new realization of $T$, say $T_W$, associated with some boundary conditions specified by the boundary unitary  $W$.

The answer to the limiting dynamics problem dates back to deep mathematical results on product formulas.
Trotter~\cite{Trotter} proved that for every $A$ and $B$ self-adjoint operators such that their sum is self-adjoint [on the intersection of their domains $D(A) \cap D(B)$] one gets
\begin{equation}
 \left( \e^{- \ii t A/N} \e^{-\ii t B /N}\right)^N   \to  \e^{-\ii t (A+B)}, \quad N \to +\infty.
\label{eq:Trotter}
\end{equation}
Unfortunately, the intersection of the domains of two self-adjoint extensions of $T$ is too small, and $T_{U_1}+ T_{U_2}$ is not self-adjoint (not even its closure!), so that this result cannot be applied to our case. 

To overcome this obstruction  it is sufficient to consider the quadratic forms associated with the operators (i.e.\ their expectation values), instead of the operators themselves. 
The expectation value of the kinetic energy  in the state $\psi$ is given by
\begin{equation}
t_{U}(\psi)=\langle\psi|T_{U}\psi\rangle_{L^2(\Omega)} = -\frac{1}{2m}\int_{\Omega} \overline{\psi(x)}  \Delta \psi(x)\; \d x .
\end{equation}
From the mathematical point of view, $t_{U}$ is the quadratic form associated with $T_{U}$, whose domain, as  discussed in subsection~\ref{quadforms}, can be considerably larger than $D(T_{U})$.

Starting from the kinetic energies $t_{U_1}$ and $t_{U_2}$, one can build the  quadratic form $t_{U_1} + t_{U_2}$ defined 
on the (dense) intersection $D = D(t_{U_1}) \cap D(t_{U_2})$. It can be proved~\cite{katobook} that if  the sum of the forms is bounded from below there is one and only one self-adjoint operator denoted by
$T_{U_1}\dot{+}T_{U_2}$, called the form sum operator of $T_{U_1}$ and $T_{U_2}$, corresponding to the form $t_{U_1} + t_{U_2}$, i.e.\ such that 
\begin{equation}
t_{T_{U_1}\dot{+}T_{U_2}}=t_{T_{U_1}}+t_{T_{U_2}}.
\end{equation}
This idea, introduced by Kato~\cite{Kato}, was elaborated by Lapidus~\cite{Lapidus1} who found the ultimate version of the Trotter product formula:
\begin{equation}
 \left( \e^{- \ii t {T_{U_1}}/N} \e^{-\ii t {T_{U_2}} /N}\right)^N   \to  \e^{-\ii t ({T_{U_1}}\dot{+}{T_{U_2}})}, \quad N \to +\infty.
\label{eq:TrotterKato}
\end{equation}
when $T_{U_1}$ and $T_{U_2}$ are bounded from below.
As a technical remark notice that, as a consequence of the weakening of the hypotheses, the convergence of the product formula when the operator sum is not self-adjoint is in a weaker topology than in Trotter's case; more precisely equation~(\ref{eq:Trotter}) holds pointwise in $t$, 
while its weaker counterpart~(\ref{eq:TrotterKato}) is valid only on average on $t$~\cite{ENZ}.
The validity of the above formula in a stronger topology is still a matter of investigation, as pointed out in~\cite{Lapidus1,ENZ}.

Summarizing, we have that
\begin{equation}
\label{eq:evollim1}
\lim_{N \to + \infty}\left(\e^{-\ii t T_{U_1}/N} \e^{-\ii  t T_{U_2}/N}\right)^N =
\e^{-\ii 2 t T_{W}},
\end{equation}
where 
\begin{equation}
T_W=\frac{T_{U_1} \;\dot{+} \; T_{U_2}}{2}.
\end{equation}
In Section~\ref{sec-cavity} we are going to prove that $T_W$ is a new realization of the operator $T$ with new boundary conditions specified by a unitary operator $W$ obtained combining $U_1$ and $U_2$ in a suitable way. The operator $W$ is given by the composition law
\begin{equation}\label{eqn:Wcomp}
W= U_1*U_2:=P_{W}+ \mathscr{C}\left(\frac{\mathscr{C}^{-1}(V_{U_1})Q_{U_1}+\mathscr{C}^{-1}(V_{U_2})Q_{U_2}}{2}\right)\,Q_W,
\end{equation}
where:
\begin{itemize}
\item  $U_{i}$ ($i=1,2$) can be uniquely decomposed into the sum $P_{U_i}+ V_{U_i}$, 
where 
$P_{U_i}$ is the spectral projection of $U_i$ on the eigenspace with eigenvalue $1$, $Q_{U_i}=I-P_{U_{i}}$, and 
$V_{U_i}=Q_{U_i}U_iQ_{U_i}$ is the restriction of $U_i$ to the range of $Q_{U_i}$;
\item $Q_W=Q_{U_1}\wedge Q_{U_2}$ is the projection onto the intersection of the ranges of $Q_{U_1}$ and $Q_{U_2}$, 
and $P_W=I-Q_W$ is the spectral projection of $W$ onto the eigenspace with eigenvalue $1$;
\item the symbols $\mathscr{C}$ and $\mathscr{C}^{-1}$ denote, respectively, the Cayley transform and its inverse. As a matter of fact the Cayley transform maps the set of self-adjoint operators onto the set of  unitary operators which do not have $1$ as an eigenvalue. In fact, for every self-adjoint operator $A$ on a Hilbert space, the operator
\begin{equation}
\mathscr{C}(A)= (A-\ii I)(A+\ii I)^{-1},
\label{eq:Cayleydef}
\end{equation}
is unitary.
Conversely, for every unitary operator $V$ such that $I-V$ is invertible, the operator
\begin{equation}
\mathscr{C}^{-1} (V)= \ii\,(I+V)(I-V)^{-1},
\end{equation}
is self-adjoint on the range of $(I-V)$.
\end{itemize}

\section{Composition law of boundary conditions for a free particle in a one-dimensional cavity} \label{sec-interval}

Before delving into the study of the composition law for boundary conditions in $\Omega \subset \R^n$, we would like to express the one-dimensional results discussed in~\cite{composition1} in a different form, suitable for a  generalization to higher dimensions.

Consider a free particle confined in the interval $\Omega = (0,1)$. Its Hamiltonian is the one-dimensional Laplacian,
\begin{equation}
T  = \frac{p^2}{2m}= - \frac{1}{2m}\frac{\d^2}{\d x^2},
\end{equation}
acting on some dense subspace of $L^2(0,1)$. Notice that in this one-dimensional case, the boundary $\partial \Omega=\{0,1\}$ consists of  two points only and the Hilbert space of the boundary values is two dimensional, that is $\mathscr{H}_b=L^2(\{0,1\})=\mathbb{C}^2$.

As proved in~\cite{aim}, the whole family of self-adjoint extensions of $T$ is in one-to-one correspondence with the possible boundary conditions coming out from $\mathrm{U}(2)$, the set of $2\times 2$ unitary matrices. More precisely, to each unitary $U \in  \mathrm{U}(2)$ there corresponds a unique self-adjoint extension 
\begin{equation}
T_U  = -\frac{1}{2m} \frac{\d^2}{\d x^2},
\end{equation}
acting on the domain 
\begin{equation}
D(T_U)=\{ \psi \in H^2(0,1):\;\ii (\I +U) \varphi = (\I-U) \dot \varphi \},
\label{eq:DTUdef}
\end{equation}
where $\varphi$ and $\dot{\varphi}$ are the boundary data of the wave function $\psi$ and are defined by
\begin{equation}
\varphi = 
\pmatrix{
\psi(0)  \cr
\psi(1)
},
\qquad \dot\varphi = \pmatrix{
-\psi'(0) \cr
\psi'(1) 
}.
\label{eq:bc21}
\end{equation}
We recall that $H^2(0,1)$ is
the Sobolev space of square integrable functions $\psi$ with square integrable first and second derivative, respectively $\psi'$ and $\psi''$.
Moreover any wave function $\psi$ in the domain of $T_U$ satisfies the  boundary conditions
\begin{equation}
\ii (\I +U) \varphi = (\I-U) \dot\varphi.
\label{eq:bc2}
\end{equation}
Notice that this parametrization of the self-adjoint extensions of $T$ in terms of boundary unitaries  differs slightly from the one presented in~\cite{composition1}; nevertheless, the two parametrizations are equivalent.

Now consider the eigenprojection of $U$ with eigenvalue $1$, from now on denoted by $P_U$. Its orthogonal projection will be denoted by $Q_U$, such that $Q_U+P_U=I$.
Notice that $P_U$ can degenerate into the identity when $U=I$, as well as it can be the zero operator when $1$ is not an eigenvalue of $U$.
 
By means of $P_U$ and $Q_U$ we are able to recast $U$ in~(\ref{eq:bc2})  into a sum on orthogonal eigenspaces, that is,
\begin{equation}
U= P_U + V_U,\qquad  V_U= U Q_U = Q_U U = Q_U\,U\,Q_U.
\end{equation}
Notice that the operator $V_U$ is unitary on the range of $Q_U$ and $1$ does not belong to its spectrum.
Having this decomposition at hand, we can give an equivalent characterization of the domain of $T_U$ by projecting equation~(\ref{eq:bc2}) onto the ranges of $P_U$ and $Q_U$:
\begin{equation}
\ii (\I +U) \varphi = (\I-U) \dot\varphi \quad \iff\quad \cases{P_U \varphi=0,\\ Q_U\dot\varphi= - K_U \varphi,}
\label{eq:projbc}
\end{equation}
where $K_U$ is (minus) the inverse Cayley transform of $V_U$, that is
\begin{equation}
K_U = - \mathscr{C}^{-1}(V_U)Q_U= - \ii\,(I+V_U)(I-V_U)^{-1}Q_{U},
\label{eq:KUdef}
\end{equation}
and $(I-V_U)^{-1}$ makes perfectly sense, since $1$ is not an eigenvalue of $V_U$.

Equation~(\ref{eq:projbc}) is valid for every unitary $U$, and particularizes to special forms according to the form of the spectral projection $P_U$.  We can distinguish among three cases:
\begin{enumerate}
\item [a)] $1$ is a double degenerate eigenvalue of $U$, therefore $P_U=I$. This corresponds to Dirichlet boundary conditions;
\item [b)] $1$ is a nondegenerate eigenvalue of $U$, therefore $U= P_U+\lambda\, Q_U$, with $|\lambda|=1$ and $\lambda \neq 1$. This case corresponds to: $Q_U\dot\varphi= \ii\, (1+\lambda)(1-\lambda)^{-1}\,Q_U \varphi\,,\,P_U\varphi=0$;
\item [c)] $1$ is not an eigenvalue of $U$, therefore $P_U=0$. For example if $U= -I$, we find Neumann boundary conditions, which are a particular case of Robin boundary conditions. The latter can be found, for example, when 
\begin{equation}
\label{robin}
U=\pmatrix{
-\e^{\ii\alpha_1} & 0 \cr
0 & -\e^{\ii\alpha_2}  
},   
\end{equation}
with $ \alpha_1,\alpha_2 \in (-\pi,\pi)$, and 
\begin{equation}
\psi'(0)= \tan \frac{\alpha_1}{2}\, \psi(0), \qquad
\psi'(1)=-\tan \frac{\alpha_2}{2}\, \psi(1).
\end{equation}
\end{enumerate}
Summing up: case a) happens when $U=I$, which corresponds to a constraint on both the values of the wave function at the boundary. Case b) provides, instead, one constraint on the values of the wave function at the boundary. For example, when $U = P_U - \e^{\ii\alpha}Q_U $, with $\alpha \in (-\pi,\pi)$ and $P_U$ being the projection onto the span of $(1,0)$, we obtain
\begin{equation}
\qquad\qquad\psi(0)=0\qquad \psi'(1)=-\tan \frac{\alpha}{2}\, \psi(1).
\end{equation}
Finally, when $1$ is not an eigenvalue, no constraint on the values of the wave function at the boundary arises from case c). It is clear now how the behavior of the wave function at the boundary is related  to the presence of the eigenvalue 1 of $U$.

\subsection{Quadratic forms} \label{sec-qforms}
In this subsection we analyze the relation between a self-adjoint operator $T_U$ and its associated quadratic form $t_U$. First of all we will explain how to get the form $t_U$ from the operator $T_U$ and then we will show how to go in the opposite way. 

Let $T_U$ be a self-adjoint extension  related to the unitary matrix $U \in \mathrm{U}(2)$.  An integration by parts yields
\begin{eqnarray*}
t_U(\psi)&=& \langle \psi\,|\, T_U\psi \rangle_{L^2(0,1)} = -\frac{1}{2m}\int_{0}^{1} \overline{\psi(x)} \;\psi''(x)\; \d x 
\nonumber\\
&=& \frac{1}{2m} \left(\int_{0}^{1} |\psi'(x)|^2 \;  \d x - \overline{\psi(1)}\; \psi'(1) + \overline{\psi(0)}\; \psi'(0)  \right)
\nonumber\\
&=&\frac{1}{2m} \left( \| \psi' \|^2_{L^2(0,1)} - \langle \varphi | \dot\varphi \rangle_{\mathbb{C}^2} \right), \quad \textrm{for all $\psi \in D(T_U)$},
\end{eqnarray*}
where $\varphi$ and $\dot\varphi$ are the boundary data~(\ref{eq:bc21}), and $ \langle \alpha | \beta \rangle_{\mathbb{C}^2}=\overline{\alpha}_1 \beta_1 + \overline{\alpha}_2 \beta_2$ is the canonical scalar product in~$\mathbb{C}^2$.
Therefore, the expectation value of the kinetic energy of the particle has both contributions from the bulk and from the boundary.
Making use of the boundary conditions  given by Eq.~(\ref{eq:projbc}), we can express the boundary contribution in a more convenient form by trading the boundary values of the derivative for the boundary values of the function. Indeed, one gets
\begin{equation}
\langle \varphi | \dot\varphi \rangle_{\mathbb{C}^2} = \langle (P_U + Q_U) \varphi |  \dot\varphi \rangle_{\mathbb{C}^2} = 
\langle \varphi | Q_U \dot\varphi \rangle_{\mathbb{C}^2} = -\langle \varphi | K_U \varphi \rangle_{\mathbb{C}^2},
\end{equation}
whence
\begin{equation}
t_U(\psi) =\frac{1}{2m} \left(\| \psi' \|^2_{L^2(0,1)} + \langle  \varphi|K_U\varphi\rangle_{\mathbb{C}^2} \right), \quad \textrm{for all $\psi \in D(T_U)$}.
\label{eq:quadraticform}
\end{equation}
Therefore, since $D(T_U)$ is a core of $t_U$, the form domain $D(t_U)$ is given by
\begin{equation}
D(t_U)=\{\psi\in H^1(0,1)\,:\,P_U\varphi=0\},
\label{eq:DtUdef}
\end{equation}
where, $H^1(0,1)$ is the Sobolev space of square integrable functions with square integrable first derivative.

The 	quadratic form $t_U$ is  closed and bounded from below, that is:
\begin{equation}
t_U(\psi)\ge -C \| \psi \|^2_{L^2(0,1)}, \qquad \textrm{for all $\psi \in D(t_U)$},
\label{eq:lowerbound}
\end{equation}
for some constant $C$ depending on the norm $\|K_U\|$ and on the continuity of the restriction map 
\begin{equation}\label{eqn:restriction_map}
\psi \in H^1(0,1) \mapsto \varphi = \psi|_{\partial \Omega}= 
\pmatrix{ 
\psi(0)  \cr
\psi(1) 
} 
\in \mathbb{C}^2.
\end{equation}
Therefore the quadratic form $t_U$ associated with a generic self-adjoint extension $T_U$ of $T$ has the following properties:
\begin{itemize}
\item the value of the form~(\ref{eq:quadraticform}) in the wave function $\psi$  is given by  the sum  of two terms:
the first one is common to all the extensions while the second one depends explicitly on the extension. Notice, indeed, that the matrix $K_U$ in~(\ref{eq:KUdef}) is, up to a sign, the inverse Cayley transform of the unitary matrix $U$ with the eigenvalue 1 stripped out;
\item the form domain $D(t_U)$ in~(\ref{eq:DtUdef}) is expressed in terms of $P_U$, the eigenprojection of $U$ with eigenvalue $1$;
\item the form $t_U$ is closed and bounded from below and its lower bound in~(\ref{eq:lowerbound}) depends on the norm of the self-adjoint matrix $K_U$ and on the continuity of the restriction map (\ref{eqn:restriction_map}).
\end{itemize}

Next, we are going to explain how to obtain the self-adjoint operator from the quadratic form. We consider a quadratic form $t$ having the same properties explained above, namely such that
\begin{equation}
\fl\quad t(\psi)=\frac{1}{2m} \left(\| \psi' \|^2_{L^2(0,1)} + \langle\varphi|K\varphi\rangle_{\mathbb{C}^2} \right),
\qquad 
D(t)=\{\psi\in H^1(0,1)\,:\,P \varphi=0\},
\label{eq:tpsidef}
\end{equation}
where $K=K^\dagger$ is a self-adjoint matrix and $P$ an orthogonal projection, such that $K P = P K = 0$. 

As above, one can show that the form $t$ is closed and bounded from below.
There is a one-to-one correspondence between the set of closed and bounded from below quadratic forms and the set of  bounded from below self-adjoint operators~\cite{katobook}, known as representation theorem.  Using this correspondence one can immediately recover the self-adjoint extension $T_U$ associated with the form $t$: 
\begin{equation}
D(T_U)=\{ \psi \in H^2(0,1):\;\ii (\I +U) \varphi = (\I-U) \dot \varphi \}
\end{equation}
where the unitary matrix $U$ is given by
\begin{equation}
U=P - \mathscr{C}(K)Q,
\label{eq:Uback}
\end{equation}
$Q=I-P$, and $\mathscr{C}$ is the Cayley transform defined in~(\ref{eq:Cayleydef}). Notice that $P$ is the eigenprojection of $U$ with eigenvalue $1$ and that $Q$ is its orthogonal projection.

\subsection{Composition law of boundary conditions} \label{sec-composition}
We now evaluate the limit of the alternating dynamics~(\ref{eq:evollim}).
As already discussed, the product formula~(\ref{eq:evollim}) holds with the form sum
$T_W=\frac{1}{2} \left(T_{U_1} \dot{+} T_{U_2} \right)$. Thus, the evaluation of the emergent dynamics in~(\ref{eq:evollim}) requires the computation of the sum
 \begin{eqnarray}
t_{12}(\psi) = \frac{t_{U_1}(\psi)+ t_{U_2}(\psi)}{2} =
\frac{1}{2m} \left(\| \psi' \|^2_{L^2(0,1)}  + \langle\varphi|K_{12}\varphi\rangle_{\mathbb{C}^2} \right),\nonumber
\nonumber \\
K_{12}=\frac{1}{2}\left(K_{U_1}+K_{U_2}\right),
\label{eq:sumforms}
\end{eqnarray}
where the form domain reads
\begin{equation}
D(t_{12}) = D(t_{U_1}) \cap D(t_{U_2})=\{\psi\in H^1(0,1)\,:\,P_{U_1}\varphi=0=P_{U_2}\varphi\}.
\end{equation}
Notice that $K_{12}$ is a self-adjoint matrix, since both $K_{U_{1}}$ and $K_{U_{2}}$ are.
Let $Q_{12}=Q_{U_1} \wedge Q_{U_2}$ be the orthogonal projection onto the intersection of the ranges of $Q_{U_1}$ and $Q_{U_2}$, that is
\begin{equation}
\mathrm{Ran} (Q_{12})= \mathrm{Ran}(Q_{U_1})\cap\mathrm{Ran}(Q_{U_2}).
\end{equation}
Moreover define 
\begin{equation}
P_{12}=I-Q_{12}.
\end{equation}
Then, the form domain of $t_{12}$ can be written in terms of the orthogonal projection $P_{12}$ as
\begin{equation}
D(t_{12})= \{\psi\in H^1(0,1)\,:\,P_{12}\; \varphi=0\}
\label{eq:Dt12def}
\end{equation}
since
\begin{equation}
N(P_{12})=\mathrm{Ran}(Q_{12})=N(P_{U_1})\cap N(P_{U_2}).
\end{equation}
Since $t_{12}$ is closed and bounded from below
by the representation theorem~\cite{katobook}, there is a unique (bounded from below) self-adjoint operator $T_{W}$ such that
\begin{eqnarray}
D(T_{W})= \{\psi\in H^{2}(0,1)\,:\,\ii (\I + W) \varphi = (\I-W) \dot\varphi\}, \nonumber \\
t_{12}(\psi)= \langle \psi, T_W \psi \rangle \qquad \textrm{for all $\psi \in D(T_W)$}.
\end{eqnarray}
Here the unitary matrix $ W \in \mathrm{U}(2)$ is given by
\begin{equation}\label{eqn:comp1}
W= P_{12} - \mathscr{C}(K_{12})\,Q_{12},
\end{equation}
and $\mathscr{C}(K_{12})$ is the Cayley transform of $K_{12}$ on the range of  $Q_{12}$. Since $P_{12}$ is the eigenprojection of $W$ with eigenvalue $1$, in accordance with our convention we have that $P_{12}=P_W$, $Q_{12}=Q_W$ and $K_{12}=K_W$. Notice that
\begin{equation}
K_{U_i}=- \mathscr{C}^{-1}(V_{U_i})Q_{U_i},\quad i=1,2,
\end{equation}
thus we obtain that~(\ref{eqn:comp1}) is exactly the formula~(\ref{eqn:Wcomp}) in the one-dimensional case.

\subsection{The path to higher dimensions}
Here we first sum up the main steps of the construction of the composition law of boundary conditions in the one-dimensional case, i.e.\ $\Omega=(0,1)$, and then we introduce the main ideas to extend the result to the general $n$-dimensional case.
\begin{enumerate}
\item[(i)] First of all we have introduced the Hilbert space of boundary values $\mathscr{H}_b=\mathbb{C}^2$, then we have defined the boundary data of a wave function
\begin{equation}
\psi \in L^2(0,1) \mapsto (\varphi, \dot{\varphi})\in \mathscr{H}_b \times \mathscr{H}_b,
\end{equation}
with $\varphi$ and $\dot{\varphi}$ being given by~(\ref{eq:bc21}),
and finally we have recalled the one-to-one correspondence~(\ref{eq:DTUdef}) between unitary operators acting on $\mathscr{H}_b$ and the self-adjoint extensions of $T$.
\item[(ii)]
We have shown that the quadratic form~(\ref{eq:quadraticform}) associated with each self-adjoint extension of $T$ is given by the sum of two terms,
the first one being independent of the extension and the second one depending on the extension through the self-adjoint matrix $K_{U}=-\mathscr{C}^{-1}(V_{U})Q_{U}$. The form domain $D(t_U)$ in~(\ref{eq:DtUdef}) is expressed in terms of the eigenprojection $P_U$ with eigenvalue $1$ of the unitary matrix $U$ specifying the extension.
Moreover the form $t_U$ is closed and bounded from below and its lower bound depends on the norm of $K_U$ and  the continuity of the restriction map~(\ref{eqn:restriction_map}).

Conversely, given a quadratic form $t$~(\ref{eq:tpsidef}), 
with $P$ being an orthogonal projection and $K$ a self-adjoint matrix, with $PK = KP =0$,
we have seen how to obtain the boundary unitary matrix $U$ by~(\ref{eq:Uback}).

\item[(iii)] Finally we have considered two different self-adjoint extensions of $T$, say $T_{U_1}$ and $T_{U_2}$, with $U_1, U_2 \in \mathrm{U}(2)$, and we have evaluated the limit of the alternating dynamics~(\ref{eq:evollim}). We have studied the sum of the quadratic forms~(\ref{eq:sumforms}),
on the domain~(\ref{eq:Dt12def}), 
where $P_{12}$ 
is the orthogonal projection  whose kernel is the intersection of the kernels of $P_{U_1}$ and $P_{U_2}$. 
Therefore, the unique self-adjoint operator $T_W$ associated with $t_{12}$ is specified by the unitary matrix $W$ obtained by a composition of $U_1$ and $U_2$:
\begin{equation}
W=U_1 \ast U_2=P_{12} - \mathscr{C}(K_{12})Q_{12}
\end{equation}
with $Q_{12}=I-P_{12}$. 
\end{enumerate}

Now we retrace this procedure and we explain step by step  the strategy to extend it to $n$ dimensions, i.e.\ to $\Omega \subset \mathbb{R}^n$. 
\begin{itemize}
\item The first difficulty is the definition of the boundary data
\begin{equation}
\psi \in L^2(\Omega) \mapsto (\varphi, \dot{\varphi}) \in \mathscr{H}_b \times \mathscr{H}_b,
\end{equation}
and the identification of the Hilbert space of boundary values $\mathscr{H}_b$. Once these aspects are clarified,  the one-to-one correspondence between the unitary operators acting on $\mathscr{H}_b$ and the self-adjoint extensions of the operator $T$ will hold as well as in the one-dimensional case. See~\cite{extensions}.
\item Given $U\in  \mathrm{U}(\mathscr{H}_b)$ we will study the quadratic form $t_U$ associated with $T_U$. Mimicking the one-dimensional case we will find that $t_U$ is given by 
\begin{equation}
t_U(\psi)= \frac{1}{2m} \left( \|\nabla \psi_{\mathrm{D}} \|_{L^2(\Omega)}^2 + \langle \varphi | K_U \varphi \rangle_{\mathscr{H}_b} \right),
\end{equation}
where $\psi_{\mathrm{D}}$ is the ``regular component" of $\psi$, which vanishes at the boundary, while $K_U$ is a self-adjoint operator defined by
\begin{equation}
K_U= - \mathscr{C}^{-1}(V_U)Q_U, \quad Q_U=I-P_U, \quad V_U=Q_U U Q_U,
\end{equation} 
and $P_U$ is the eigenprojection of $U$ with eigenvalue $1$. Understanding  the meaning of $\psi_{\mathrm{D}}$ is one of the main  difficulties in the study of the quadratic form $t_U$. A further complication arises from the fact that, in the $n$-dimensional case, the operator $K_U$ can be unbounded. This means that the form $t_U$ is not, in general, bounded from below. In order to avoid this problem, we will consider only unitary operators $U \in \mathrm{U}(\mathscr{H}_b)$  having a spectrum with a gap around the point $1$: this condition ensures that $K_U$ is bounded.  
\item Finally, we will extend the composition law for boundary conditions  assuming that the spectra of $U_1, U_2 \in \mathrm{U}(\mathscr{H}_b)$ are both gapped around the point $1$. This compatibility condition is necessary to extend the construction of the composition $U_1 \ast U_2$ to the $n$-dimensional case because it ensures that the form sum $t_{12}$ is densely defined, closed and bounded from below, and allows us to identify the self-adjoint extension corresponding to $t_{12}$.
\end{itemize}

\section{A composition law for a free particle in a cavity} \label{sec-cavity}
In this section we are going to extend the composition law determined in subsection~\ref{sec-composition}. Since we are going to describe a free particle in a cavity $\Omega$ we need some results about self-adjoint extensions of the operator $T$ in terms of boundary conditions. This was largely discussed in~\cite{grubb68,aim,extensions}.
\subsection{The framework}
Let $\mathrm{\Omega}$ be an open bounded set in $\mathbb{R}^n$, whose boundary is regular. Let $\nu$ be the normal to $\partial\Omega$, by convention $\nu$ is oriented towards the exterior of $\Omega$. We define for a regular function $\psi$ its normal derivative along $\nu$:
\begin{equation}  
\partial_\nu\psi=\nu\cdot\left(\nabla\psi\right)|_{\partial\Omega}.
\end{equation}
Notice that, by definition, $\partial_\nu\psi$ is a function settled on $\partial\Omega$.

We consider the operator
\begin{equation}
T=- \frac{1}{2m}\Delta,
\end{equation}
defined on  some dense subset of $L^2(\Omega)$, for example $\mathscr{D}(\Omega)$, the space of test functions in $\Omega$ (compactly supported and smooth functions in $\Omega$).
Wave functions on this space are not able to provide information about the boundary because for every $\psi\in\mathscr{D}(\Omega)$: $\psi|_{\partial\Omega}=\partial_\nu\psi=0$.
Let $T^\dagger$ be the adjoint  of $T$ with domian 
\begin{equation}
\label{eq:DT*}
D(T^\dagger)=\{\psi\in L^{2}(\Omega)\,:\, \Delta 	\psi \in L^2(\Omega)\}.
\end{equation}
We are going to identify the Hilbert space of the boundary values $\mathscr{H}_b$. It can be proved~\cite{lm} that  the ``restriction" to the boundary $\partial \Omega$ of  a wave function $\psi \in L^2(\Omega)$,  denoted by $\psi|_{\partial\Omega}$, belongs to $H^{-\frac{1}{2}}(\partial \mathrm\Omega)$,  the fractional Sobolev space of order $-1/2$. 
Therefore in the $n$-dimensional case the Hilbert space of boundary values  $\mathscr{H}_b=H^{-\frac{1}{2}}(\partial \mathrm\Omega)$ is infinite dimensional, unlike the one-dimensional case.

We are now ready to state the following result, proved in~\cite{extensions}, about the self-adjoint extensions of $T$.
\begin{theorem}\label{thm:GG}
The set of all self-adjoint extensions of T is 
\begin{equation}
\left\{T_U: D(T_U) \to L^2(\mathrm\Omega), \, U \in \mathrm{U}(\mathscr{H}_b) \right\},
\end{equation}
where for all $U \in \mathrm{U}(\mathscr{H}_b)$
\begin{equation}
D(T_U)=\left\{\psi \in D(T^\dagger)\,:\, i (I+U)\varphi =(I-U)\, \dot\varphi \,\right\},
\end{equation}
and the boundary data $(\varphi,\dot\varphi)$ are specified in Eq.~(\ref{eq:boundarydata}).
\end{theorem}

In the one-dimensional case, equation~(\ref{eq:bc2}) provided a thorough parametrization of the self-adjoint extensions of the operator $T$. Theorem~\ref{thm:GG} states that Equation~(\ref{eq:bc2}) is still valid for a general $\Omega\subset\mathbb{R}^n$ through a suitable identification of the boundary data $(\varphi,\dot\varphi)$. In what follows we would like to provide the reader with explanations about the boundary data $(\varphi,\dot\varphi)$.

As discussed above, the restriction $\psi|_{\partial\Omega}$ of a wave function $\psi \in L^2(\Omega)$ to the boundary $\partial \Omega$ belongs to $\mathscr{H}_b=H^{-\frac{1}{2}}(\partial \mathrm\Omega)$, while its normal derivative $\partial_\nu\psi$ belongs to a more irregular space (more precisely $\partial_\nu\psi$ belongs to $H^{-\frac{3}{2}}(\partial \mathrm\Omega)$)~\cite{lm}.

For this reason equation~(\ref{eq:bc2}) cannot hold by naively interpreting the boundary data $(\varphi,\dot\varphi)$ as the pair $(\psi|_{\partial\Omega},\partial_\nu\psi)$. Indeed, the elements of the pair are settled on different Hilbert spaces
and the existence of an operator $U$ acting both on $\psi|_{\partial\Omega}$ and $\partial_\nu\psi$  becomes meaningless.
Moreover, the boundary values $\psi|_{\partial\Omega}$ and $\partial_\nu\psi$ are \emph{not} independent data, as discussed in~\ref{sec:appendixA}, and
one can show that only the normal derivative of a ``regular'' component $\psi_{\mathrm{D}}$ of $\psi$  is 
independent of $\psi|_{\partial\Omega}$.

In order to define the regular component $\psi_{\mathrm{D}}$ of $\psi$, we need a useful decomposition of the domain of the adjoint $D(T^\dagger)$ given in~(\ref{eq:DT*}):
\begin{equation}\label{eqn:dec}
D(T^\dagger)=D(T_{\mathrm{D}})+N(T^\dagger), \qquad \psi = \psi_{\mathrm{D}} + \psi_0,
\end{equation}
where $T_{\mathrm{D}}$ is the self-adjoint extension of $T$ with Dirichlet boundary conditions, that is 
on the domain
\begin{equation}
D(T_{\mathrm{D}})=\{\psi \in H^{2}(\mathrm\Omega)\,:\, \psi|_{\partial\Omega}=0 \},
\end{equation}
and
\begin{equation}
N(T^\dagger)=\{\psi \in D(T^\dagger)\,:\, \Delta \psi =0 \} \,
\end{equation}
is the kernel of $T^\dagger$.
In other words, every $\psi \in D(T^\dagger)$ can be uniquely decomposed in the sum $\psi_{\mathrm{D}}+\psi_0$, where $\psi_{\mathrm{D}}\in  D(T_{\mathrm{D}})$ is a function vanishing on the boundary, $\psi_{\mathrm{D}}|_{\partial\Omega}=0$, and $\psi_0$ is a harmonic function, $\Delta\psi_0=0$. 
See~\ref{sec:appendixA}  for more details.

We are finally in the right position to define the boundary data $(\varphi,\dot\varphi)$ of a wave function $\psi \in D(T^\dagger)$:
\begin{equation} \label{eq:boundarydata}
\varphi=\psi|_{\partial\Omega} = \psi_0|_{\partial\Omega}, \qquad \dot\varphi=\Lambda\, \partial_{\nu}\psi_{\mathrm{D}} ,
\end{equation}
where $\psi_{\mathrm{D}}$ is the regular component of $\psi$ in the sense of the decomposition~(\ref{eqn:dec}). Here $\Lambda=(I-\Delta_{\partial\Omega})^{\frac{1}{2}}$, where $\Delta_{\partial\Omega}$ is the Laplace-Beltrami operator on $\partial\Omega$~\cite{berger}, and its role is merely to pull back $\partial_{\nu}\psi_{\mathrm{D}} \in H^{\frac{1}{2}}(\partial\Omega)$  to $\mathscr{H}_b=H^{-\frac{1}{2}}(\partial\Omega)$, the common Hilbert space of the boundary data. Any unitary map from $H^{\frac{1}{2}}(\partial\Omega)$ to $H^{-\frac{1}{2}}(\partial\Omega)$ will do, and the reader can safely ignore its presence henceforth.

For further mathematical details see~\cite{extensions}. 

\subsection{Parametrization of the self-adjoint extensions of $T$ by means of spectral projections}

It is of interest to recast the former parametrization of $T_U$ by  isolating the contribution of the eigenvalue 1 from the spectrum of $U$ and expressing it in terms of spectral projections.

In analogy with subsection~\ref{sec-interval} we define $P_U$ as the eigenprojection of $U$ with eigenvalue $1$ and $Q_U=I-P_U$ its orthogonal projection.

\begin{figure}[tbp]
\centering
\includegraphics[width=0.5\columnwidth]{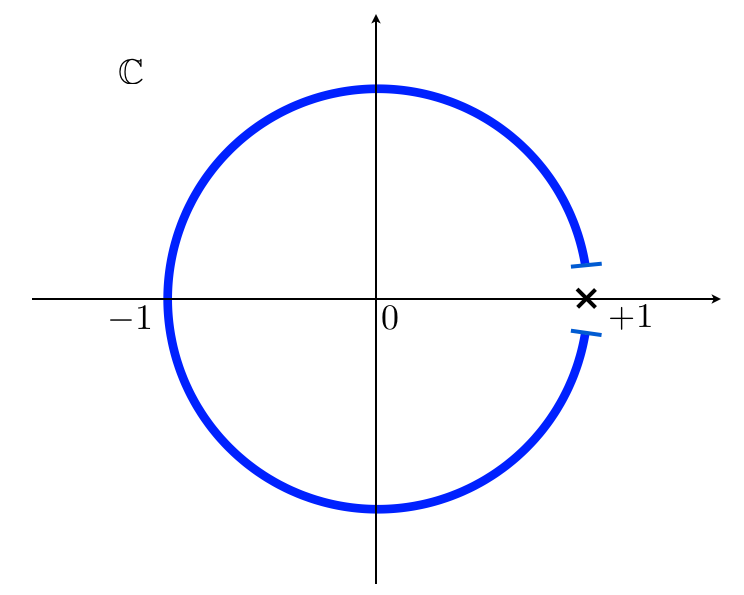}
\caption{The spectrum of $U$ has a gap around the point 1 in the complex plane.}
\label{fig:gapped}
\end{figure}

Then, the unitary operator $U$ can be decomposed in the sum $U=P_U + V_U$, where $V_U= Q_U\,U\,Q_U$.
After projecting onto the two mutually orthogonal subspaces given by $P_U$ and $Q_U$, the domain of $T_{U}$ reads
\begin{equation}
\label{eqbcdelta}
D(T_U)=\left\{\psi \in D(T^\dagger):\,P_U\varphi=0\,,\,Q_U\dot\varphi= - K_U \varphi \right\},
\end{equation}
where $K_U$ is given by (minus) the Cayley transform of $V_U$,
\begin{equation}
\label{eq:cay}
K_U = - \mathscr{C}^{-1}(V_U)Q_U= - \ii\,(I+V_U)(I-V_U)^{-1}Q_{U}.
\end{equation}
Even in higher dimensions  the operator $(I-V_U)$ is invertible, since the eigenvalue 1 was stripped out, but it may be unbounded. The operator $(I-V_U)^{-1}$ will be bounded as long as 1 does not belong to the spectrum of $V_U$, that is as long as 
the spectrum of $U$ has a gap around the point 1 (Figure~\ref{fig:gapped}). For this reason, from now on, we are going to make the latter our working assumption~\cite{juanma}.

\subsection{Quadratic forms}\label{quadforms}

As already sketched for the one-dimensional case we provide an explicit expression for the expectation value of the kinetic energy of a free particle in a cavity $\Omega$. 

We are going to prove that the expectation value of the kinetic energy $T_U$ is
\begin{equation}
t_U(\psi)= \frac{1}{2m} \left( \|\nabla \psi_{\mathrm{D}} \|^2_{L^2(\Omega)} + \langle \varphi|K_U\varphi\rangle_{\mathscr{H}_b} \right), 
\label{eq:tUndim}
\end{equation}
where $\psi_{\mathrm{D}}$ is the regular component of $\psi$ in the sense of  the decomposition~(\ref{eqn:dec}), $K_U$ is the operator defined in~(\ref{eq:cay}) and $\mathscr{H}_b =H^{-\frac{1}{2}}(\partial \Omega)$.

Take $\psi\in D(T_U)\cap C^{\infty}(\overline \Omega)$, since $D(T_U)\subset D(T^\dagger)=D(T_{\mathrm{D}})+N(T^\dagger)$, it follows that it can be decomposed as $\psi =\psi_{\mathrm{D}} + \psi_0$. Therefore,
\begin{eqnarray}
2m \, t_U(\psi)&=& - \langle\psi\,|\Delta \psi\rangle_{L^2(\Omega)}= - \langle\psi\,|\Delta \psi_{\mathrm{D}}\rangle_{L^2(\Omega)} - \langle\psi\,|\Delta \psi_0\rangle_{L^2(\Omega)} \nonumber \\
&=& - \langle\psi\,|\Delta \psi_{\mathrm{D}}\rangle_{L^2(\Omega)}= - \langle\psi_{\mathrm{D}}\,|\Delta \psi_{\mathrm{D}}\rangle_{L^2(\Omega)} -\langle\psi_0\,|\Delta \psi_{\mathrm{D}}\rangle_{L^2(\Omega)} ,
\end{eqnarray}
where we used twice the decomposition of $\psi$ and once the property $\Delta \psi_0=0$.
Next by means of the Gauss-Green formulas 
we will show that
\begin{enumerate}
\item [(i)] $\langle\psi_{\mathrm{D}}\,|\Delta \psi_{\mathrm{D}}\rangle_{L^2(\Omega)}=- \|\nabla \psi_{\mathrm{D}} \|^2_{L^2(\Omega)}$;
\item [(ii)] $\langle\psi_0\,| \Delta \psi_{\mathrm{D}}\rangle_{L^2(\Omega)}= \langle \psi
|\partial_\nu\psi_{\mathrm{D}}\rangle_{L^2(\partial \Omega)}$;
\end{enumerate}
so that, by putting them  together, we obtain
\begin{equation}
2m \, t_{U}(\psi)=\|\nabla \psi_{\mathrm{D}} \|^2_{L^2(\Omega)}-\langle \psi 
|\partial_\nu\psi_{\mathrm{D}}\rangle_{L^2(\partial \Omega)}.
\end{equation} 
Let us begin from (i):
\begin{eqnarray}
\fl \quad   \langle\psi_{\mathrm{D}}\,|\Delta \psi_{\mathrm{D}}\rangle_{L^2(\Omega)}&=& \int_{\Omega}\overline{\psi_{\mathrm{D}}(x)}\;\Delta\psi_{\mathrm{D}}(x) \;\d x
= - \int_{\Omega}|\nabla\psi_{\mathrm{D}}(x)|^2\;\d x + \int_{\partial\Omega}\overline{\psi_{\mathrm{D}}}\;\partial_\nu\psi_{\mathrm{D}}\;\d S \nonumber \\
&=& - \int_{\Omega}|\nabla\psi_{\mathrm{D}}(x)|^2\;\d x= - \|\nabla \psi_{\mathrm{D}} \|^2_{L^2(\Omega)},
\end{eqnarray}
where we used the property $\psi_{\mathrm{D}}|_{\partial\Omega}=0$.

Next we move on to the computation of (ii). Notice first that
\begin{equation}
\fl\qquad  \int_{\Omega}\overline{\nabla\psi_{\mathrm{D}}(x)}\cdot\nabla\psi_0(x)\;\d x 
= -\int_{\Omega}\overline{\psi_{\mathrm{D}}(x)}\,\Delta\psi_0(x)\;\d x + \int_{\partial\Omega}\overline{\psi_{\mathrm{D}}} \; \partial_\nu\psi_0\;\d S
=0 ,
\label{eq:orth}
\end{equation}
since $\psi_{\mathrm{D}}|_{\partial\Omega}=0$ and $\Delta \psi_0=0$.
Therefore,
\begin{eqnarray}
\fl \langle\psi_0\,|\Delta \psi_{\mathrm{D}}\rangle_{L^2(\Omega)}&=&\int_{\Omega}\overline{\psi_0(x)}\;\Delta\psi_{\mathrm{D}}(x) \;\d x = -\int_{\Omega}\overline{\nabla\psi_0(x)}\cdot\nabla\psi_{\mathrm{D}}(x)\;\d x +\int_{\partial\Omega}\overline{\psi_0}\; \partial_\nu\psi_{\mathrm{D}}\;\d S \nonumber \\
&=& \int_{\partial\Omega} \overline{\psi} 
\;\partial_\nu\psi_{\mathrm{D}}\;\d S= \langle \psi  
|\partial_\nu\psi_{\mathrm{D}}\rangle_{L^2(\partial \Omega)} ,
\end{eqnarray}
where we made use of equation~(\ref{eq:orth}) and the equality $\psi_0|_{\partial\Omega}=\psi|_{\partial\Omega}$.

Putting together all the ingredients we find that, for all $\psi\in D(T_U)\cap C^{\infty}(\overline \Omega)$,
\begin{eqnarray}
2 m \, t_U(\psi)
&=&\|\nabla \psi_{\mathrm{D}} \|^2_{L^2(\Omega)} - \langle \psi|\partial_\nu\psi_{\mathrm{D}}\rangle_{L^2(\partial \Omega)}
\label{eqn2} \nonumber\\
&=&\|\nabla \psi_{\mathrm{D}} \|^2_{L^2(\Omega)}-\langle \varphi|\dot\varphi\rangle_{\mathscr{H}_b}\label{eqn4} 
=\|\nabla \psi_{\mathrm{D}} \|^2_{L^2(\Omega)} + \langle \varphi|K_U\varphi\rangle_{\mathscr{H}_b}\label{eqn5}.
\end{eqnarray}
where  we used 
the definition of the boundary values $(\varphi, \dot{\varphi})$ in~(\ref{eq:boundarydata}) and
the boundary conditions in~(\ref{eqbcdelta}). By a density argument one finally gets~(\ref{eq:tUndim})
for all $ \psi \in D(T_U)$.

The mathematical expression of  the kinetic energy $t_U$ of a free particle in a cavity $\Omega$ highly resembles the one given in equation~(\ref{eq:quadraticform}) for the one-dimensional case. Though this apparent similarity, the two equations differ considerably. For example the contribution from the bulk is due to the regular part of $\psi$, that is $\psi_{\mathrm{D}}$, rather than from the whole function. Moreover the boundary term is in general unbounded. It is bounded as long as $K_{U}$ is bounded, which is guaranteed by the condition of~$U$ with gapped spectrum.

Finally, since $D(T_U)$ is a core of $t_U$ we find that
\begin{equation}
D(t_U)=\{\psi\in H^1_0(\Omega)+N(T^\dagger)\,:\,P_U \varphi=0\},
\end{equation}
where  $H^1_0(\Omega)$ is the subspace of $H^1(\Omega)$ whose functions vanish at the boundary. For more details see~\cite{extensions}.

\subsection{Composition law}
We now evaluate the limit of the alternating dynamics~(\ref{eq:evollim1}) in the case of a free particle confined in a cavity $\Omega\subset\mathbb{R}^n$.
Once more the product formula~(\ref{eq:evollim1}) holds with the form sum
\begin{equation}
T_{W} = \frac{T_{U_1} \; \dot{+} \; T_{U_2}}{2}.
\end{equation}
Following the one-dimensional case we carry on the computation of the sum
\begin{eqnarray}
t_{12} (\psi)= \frac{t_{U_1}(\psi) + t_{U_2}(\psi)}{2}= \frac{1}{2m} \left( \| \nabla\psi_{\mathrm{D}} \|^2_{L^2(\Omega)} + \langle\varphi|K_{12}\varphi\rangle_{\mathscr{H}_b} \right),\nonumber
\\
K_{12}=\frac{1}{2}\left( K_{U_1}+K_{U_2}\right),
\label{eq:sumndim}
\end{eqnarray}
and its domain
\begin{equation}
D(t_{12}) = \{\psi\in H^1_0(\Omega)+N(T^\dagger)\,:\,P_{12}\varphi=0\}.
\end{equation}
We stress that $K_{12}$ is a bounded self-adjoint operator, since both $K_{U_1}$ and $K_{U_2}$ are, and thus the quadratic form $t_{12}$ is closed and bounded from below. Therefore, by  the representation theorem~\cite{katobook}, there exists a unique self-adjoint extension $T_W$ of $T$ such that
\begin{eqnarray}
t_{12}(\psi)= \langle \psi| T_W \psi \rangle_{L^(\Omega)}, \qquad \textrm{for all $\psi \in D(T_W)$}, 
\end{eqnarray}
where 
\begin{equation}
D(T_{W})= \{\psi\in D(T^\dagger)\,:\,\ii (\I + W) \varphi = (\I-W) \dot{\varphi}\}.
\end{equation}
We stress that $P_{W}=P_{12}$, so that $W$ can be explicitly built from $P_{12}$ and $K_{12}$ as
\begin{equation}
W= P_{12} - \mathscr{C}(K_{12})Q_{12},
\end{equation}
where $\mathscr{C}(K_{12})$ is the Cayley transform of $K_{12}$ on the range of $Q_{12}$. This complete our proof of Eq.~(\ref{eqn:Wcomp}). 

\section{Conclusions} \label{sec-concl}
We have studied the dynamics of a quantum particle confined in a cavity subject to two rapidly alternating boundary conditions. The limit dynamics has been expressed in terms of new boundary conditions, which have emerged from a dynamical composition law.

Differences and similarities between the one-dimensional and the $n$-dimensional case have been shown and largely discussed. While in the one-dimensional case the composition law is expressed in terms of unitary matrices, in the $n$-dimensional case, instead, it is obtained through unitary operators on an infinite dimensional Hilbert space. For this reason the spectral properties of the operators have to be taken into consideration and handled with care.

In the end, in the near future it would be interesting both to apply the aforementioned results to cavities having interesting geometrical properties and to analyze the case of the Dirac operator. Further investigations will involve concrete analysis and applications such as the Casimir effect or the quantum Hall effect. 

\appendix

\section{A useful decomposition}
\label{sec:appendixA}
In this appendix we would like to provide the reader with  further details about the decomposition of the domain~(\ref{eq:DT*}),
\begin{equation}
D(T^\dagger)=D(T_{\mathrm{D}})+N(T^\dagger),
\label{eq:dirctsum}
\end{equation} 
where
\begin{eqnarray}
\fl \qquad D(T_{\mathrm{D}})=\{\psi \in H^{2}(\mathrm\Omega)\,:\, \psi|_{\partial\Omega}=0\}, \qquad
N(T^\dagger) = \{\psi \in L^2(\Omega) \,:\, \Delta 	\psi=0\}.
\end{eqnarray}
As explained in Section~\ref{sec-cavity}, this decomposition is crucial to define the regularized normal derivative of a wave function and its boundary data $(\varphi, \dot{\varphi})$ in~(\ref{eq:boundarydata}). Moreover, it enters in the definition of the kinetic energy~(\ref{eq:tUndim}).

In order to understand the meaning of this decomposition, we  explain here how to decompose a smooth  function $\psi\in C^{\infty}(\overline{\Omega}) \subset D(T^\dagger)$ into the sum $\psi=\psi_{\mathrm{D}}+\psi_0$, with $\psi_{\mathrm{D}}|_{\partial \Omega}=0$ and $\Delta \psi_0=0$. First of all we define $g = \psi |_{\partial \Omega}$ and solve the (boundary value) electrostatic problem:
\begin{equation}\label{eq:DirSisthom}
\cases{-\Delta u =0 & on $\Omega$\\
\quad\,\,\,\,u=g & in $\partial\Omega$}.
\end{equation}
The solution $\psi_0$ of~(\ref{eq:DirSisthom}) represents the electrostatic potential in the cavity $\Omega$ with the given value 
$g=\psi |_{\partial \Omega}$ on the boundary. Next, we define $\psi_{\mathrm{D}}$ as $\psi-\psi_0$; manifestly $\psi_{\mathrm{D}}|_{\partial\Omega}=0$. Therefore we can write $\psi=\psi_{\mathrm{D}}+\psi_0$ with $\psi_{\mathrm{D}}|_{\partial\Omega}=0$ and $\Delta \psi_0=0$.
This decomposition can be extended by a density argument~\cite{grubbook} to $D(T^\dagger)$,
obtaining~(\ref{eq:dirctsum}).

Notice that $D(T_{\mathrm{D}})$ represents the domain of a self-adjoint extension of $T$ (the one specified by Dirichlet boundary conditions) and it is made up by much more regular functions than $N(T^\dagger)$.
In fact, the space $N(T^\dagger)$ contains functions that can be very irregular on $\partial \Omega$. This is a fairly interesting phenomenon in potential theory: a harmonic functions, which is extremely regular (analytic) in the interior of $\Omega$ can become very irregular on its boundary.  

For example, consider the function $f(x,y)=1/z = 1/ (x+\ii y)$ in the upper half-plane $\Omega=\{(x,y) \in \mathbb{R}^2:y >0\}$.  
It is evident that though being harmonic on $\Omega$, the function $f$ diverges at the point $(0,0) \in  \partial\Omega$. In fact, its boundary value is a distribution! Indeed one gets
\begin{equation}
\lim_{y\downarrow0}f(x,y)=P.V.\frac{1}{x}-\ii\pi\delta(x),
\end{equation}
where $P.V.$ is the Cauchy principal value and $\delta$ the Dirac delta.

\section{Semi-gapped boundary unitaries}
\label{sec:appendixB}

\begin{figure}[tbp]
\centering
\includegraphics[width=0.5\columnwidth]{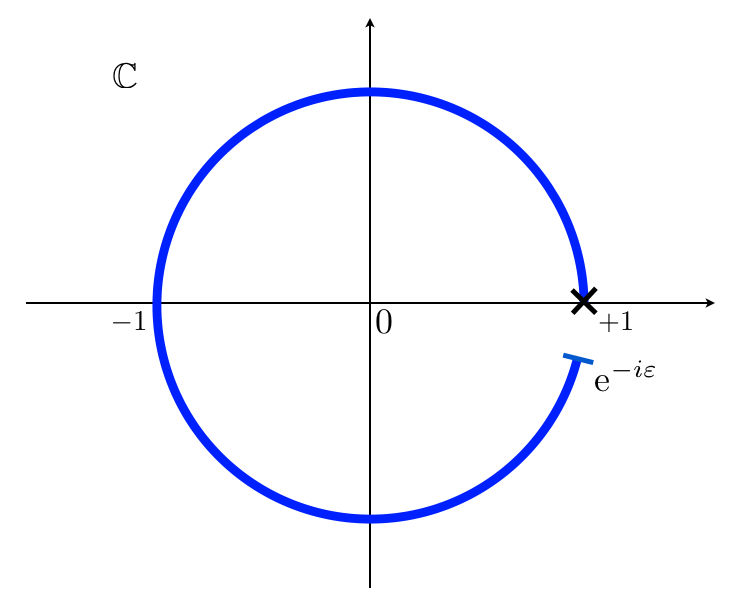}
\caption{If $K_U$ is  bounded from below 
then the spectrum of $V_U$ has a gap in the spectrum just below the point 1. 
}
\label{fig:gapped2}
\end{figure}

In this appendix we consider the extension of the composition law~(\ref{eqn:Wcomp}) to boundary unitaries $U_i$ which are not gapped, and thus to unbounded boundary operators $K_{U_i}$.
In general the sum of two unbounded (self-adjoint) operators $K_{U_1} + K_{U_2}$ is a touchy business. However, if the two operators are bounded from below, the situation can be kept somewhat under control.

It can be proved~\cite{grubb74} that a self-adjoint extension  $T_U$ of $T$ is bounded from below as long as the corresponding operator $K_U$, which appears in equation~(\ref{eq:cay}), is bounded from below. 
In turn, by the properties of the Cayley transform~(\ref{eq:Cayleydef}), it is easy to see that $K_U$ is bounded from below if and only the spectrum of $U$  has a gap just below the point $1$, namely the set $\{\e^{\ii \alpha} \,:\, \alpha\in(-\varepsilon, 0) \}$ belongs to the resolvent set of $U$ for some $\varepsilon>0$. We will call such a $U$ semi-gapped. See figure~\ref{fig:gapped2}, where $\beta = - \cot(\varepsilon/2)$.

Suppose then that $T_{U_1}$ and $T_{U_2}$ are bounded below self-adjoint operators. We remind the reader that the hypothesis of boundedness from below is fundamental for the Trotter-Kato formula~(\ref{eq:TrotterKato}) to hold.
Next, consider the sum of their quadratic forms
\begin{eqnarray}
t_{12} (\psi)= \frac{t_{U_1}(\psi) + t_{U_2}(\psi)}{2}= \frac{1}{2m} \left(\| \nabla\psi_{\mathrm{D}} \|^2_{L^2(\Omega)} + \langle\varphi|K_{12}\varphi\rangle_{\mathscr{H}_b} \right),\nonumber
\\
K_{12}=\frac{1}{2}\left( K_{U_1}+K_{U_2}\right), \qquad D(K_{12})=D(K_1)\cap D(K_2).
\end{eqnarray}
It is defined on the domain
\begin{eqnarray}
\fl\qquad  D(t_{12}) = D(t_{U_1})\cap D(t_{U_2}) = \{\psi\in H^1_0(\Omega)+N(T^\dagger)\,:\,P_{12}\varphi=0\}\cap D_{K_{12}},
\end{eqnarray}
where
\begin{equation}
D_{K_{12}}=\{\psi\in L^2(\Omega):\psi|_{\partial\Omega}\in D(K_{12})\subset \mathscr{H}_b \}.
\end{equation}

Now suppose that $K_{12}$ is self-adjoint on $N(P_{12})$. It may happen, in fact, that the sum $K_1 + K_2$ is not self-adjoint, and $D(K_1)\cap D(K_2)$ could even reduce to the trivial space! In particular, it is sufficient, for example, that either $K_1$ or $K_2$ is bounded, so that their sum is surely  self-adjoint. The latter situation happens when, for example,  
1 is in the resolvent set of $V_{U_1}$, that is when $U_1$ is gapped.

Then, by the representation theorem in~\cite{extensions} the form $t_{12}$ is the expectation value of a self-adjoint operator, $T_W$ ($W$ being a unitary operator on $\mathscr{H}_b$), which is (half) the form sum operator of $T_{U_1}$ and $T_{U_2}$, namely $T_{W} = (T_{U_1} \; \dot{+} \; T_{U_2})/2$. Notice that, by the previous discussion it follows that $T_W$ is  bounded below  because $K_{12}$ is.

\ack 
This work was  supported by Cohesion and Development Fund 2007-2013 - APQ Research Puglia Region ``Regional program supporting smart specialization and social and environmental sustainability - FutureInResearch'', by the Italian National Group of Mathematical Physics (GNFM-INdAM, Progetto Giovani), and by Istituto Nazionale di Fisica Nucleare (INFN) through the project ``QUANTUM''.

\end{document}